\documentclass[12pt,preprint]{emulateapj}

\newcommand{\vv}[1]{\textbf{#1}}	

\newcommand{\Msun}{M_{\odot}}
\newcommand{\Ma}{M_{\rm a}}
\newcommand{\Mc}{M_{\rm c}}

\begin{document}
\title
{
Magnetic burial and the harmonic
content of millisecond oscillations in thermonuclear
X-ray bursts
}
\author
{D. J. B. Payne and A. Melatos}
\affil
{School of Physics, University of Melbourne,
Parkville, VIC 3010, Australia.
}
\email
{dpayne@physics.unimelb.edu.au}
\begin{abstract}
Matter accreting onto the magnetic poles of a neutron star
spreads under gravity towards the magnetic equator,
burying the polar magnetic field and compressing it into
a narrow equatorial belt.
Steady-state, Grad-Shafranov calculations with a self-consistent
mass-flux distribution
(and a semi-quantitative treatment of Ohmic diffusion)
show that, for
$\Ma \gtrsim 10^{-5}\Msun$,
the maximum field strength and latitudinal half-width of the
equatorial magnetic belt are
$B_{\rm max} = 5.6\times 10^{15} (\Ma/10^{-4}\Msun)^{0.32}$ G and
$\Delta\theta = \max[3^{\circ} (\Ma/10^{-4}\Msun)^{-1.5},3^{\circ}
(\Ma/10^{-4}\Msun)^{0.5}(\dot{M}_{\rm a}/10^{-8}\Msun {\rm yr}^{-1})^{-0.5}]$
respectively,
where $\Ma$ is the total accreted mass and $\dot{M}_{\rm a}$ is the
accretion rate.
It is shown that the belt prevents north-south
heat transport by conduction, convection, radiation, and
ageostrophic shear.
This may explain why millisecond 
oscillations observed in the tails of thermonuclear
(type I) X-ray bursts in low-mass X-ray binaries are highly sinusoidal:
the thermonuclear flame is sequestered in the magnetic hemisphere
which ignites first.
The model is also consistent with the occasional occurrence of closely
spaced pairs of bursts.
Time-dependent, ideal-magnetohydrodynamic simulations confirm that
the equatorial belt is not disrupted by
Parker and interchange instabilities.
\end{abstract}


\section{Introduction}
\label{sec:acc1}

Thermonuclear (type I) X-ray bursts 
are observed from $70$ of the $160$
low-mass X-ray binaries (LMXBs)
discovered to date
\citep{str03}.
They recur every few hours to days,
when the accreted surface layer of the neutron star ignites and
is incinerated by
hydrogen and helium burning.
Brightness oscillations with millisecond periods are observed during
thermonuclear X-ray bursts in 13 LMXBs
\citep{mun02,bil04}.
They arise during burst onset because
the stellar photosphere is temporarily patchy while the thermonuclear flame
spreads from its ignition point to cover the star.

A surprising property of the burst oscillations is that they often
persist into the tails of bursts, after the flame is expected to
have engulfed the star
\citep{str97}.
Equally surprising is how
sinusoidal the oscillations are.
\citet{mun02} analyzed the harmonic content of 59 oscillations
from six sources, observed with
the \emph{Rossi X-Ray Timing Explorer} (\emph{RXTE}),
and found that the Fourier amplitudes of
integer and half-integer
harmonics are less than 5 and 10 per cent of the maximum signal
respectively.
These data imply that
if there is one hot spot on the surface, it must lie near the
rotational pole or cover an entire hemisphere,
whereas
if there are two, antipodal hot spots, they must lie near
the rotational equator
\citep{mun02}.
Most theories that explain why the oscillations persist into the
tails of the bursts, e.g.
uneven heating/cooling during photospheric uplift
\citep{str97}, or cyclones driven by zonal shear in a geostrophic flow
\citep{spi02}, are hard pressed to account for the pattern of hot spots
implied by the Fourier data.
Global $r$-modes in the neutron star ocean can
divide the photosphere into symmetric
halves \citep{hey04},
but the physical mechanism converting $r$-mode
density perturbations to brightness
oscillations is unclear \citep{mun02}.
\citet{cum05} showed that differential rotation between the pole
and equator of $\gtrsim 2$ per cent can also excite unstable modes.

In this paper,
we show that an equatorial belt of
intense magnetic field, compressed by accreted material spreading
away from the magnetic poles, can impede
thermal transport between the hemispheres
of the star.
In \S 2, we review the physics of magnetic burial
and compute the maximum field strength and width of
the equatorial magnetic barrier.
In \S 3, we estimate the
efficiency of thermal transport across the barrier by
conduction, convection, radiation, and
ageostrophic shear,
and investigate whether cyclonic flows can disrupt the
barrier.
The implications for the harmonic content of 
burst oscillations are explored in \S 4.

\section{Magnetic burial}

\subsection{Grad-Shafranov equilibria}
\label{sec:gradshaf}
\begin{figure}
\centering
\leavevmode
\includegraphics[height=65mm]{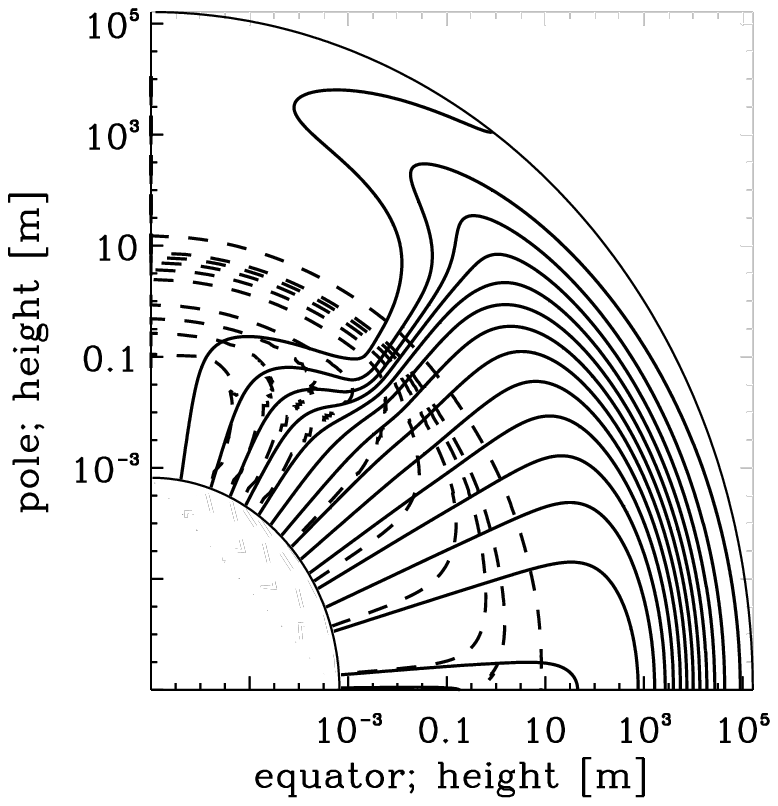} \\
\includegraphics[height=65mm]{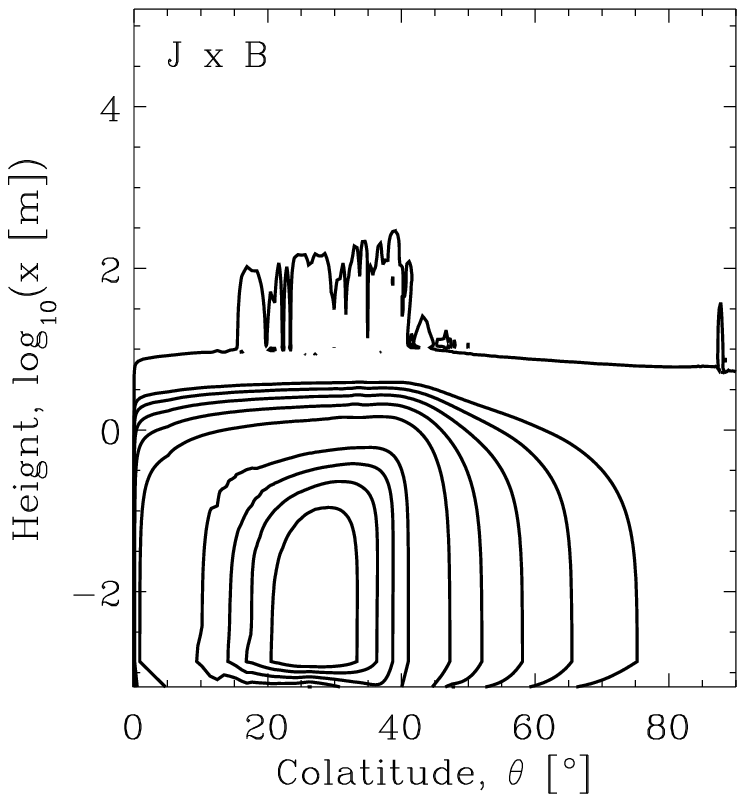} \\
\includegraphics[height=65mm]{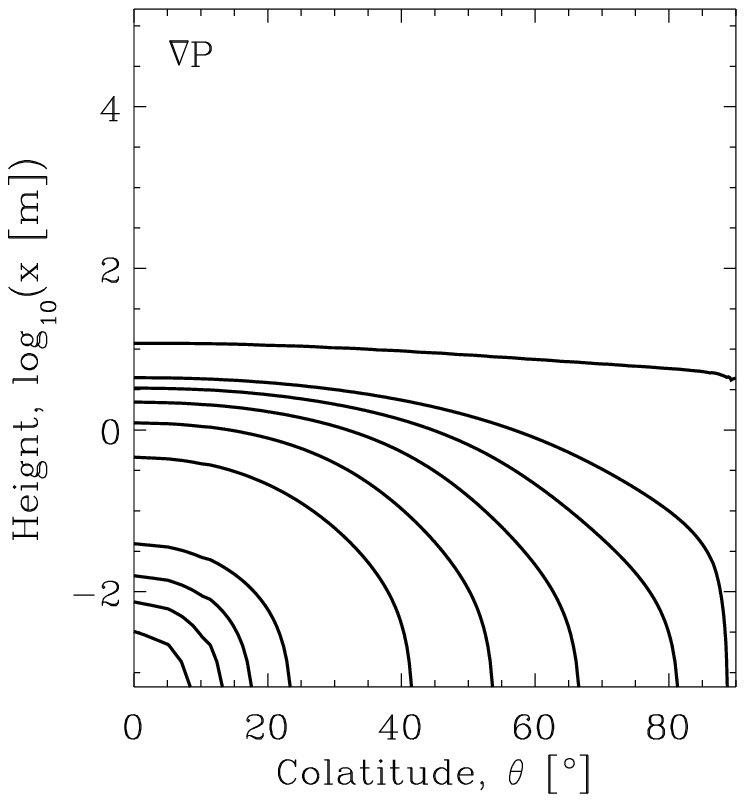} \\
\caption{
(\emph{top}) Equilibrium magnetic field lines (solid curves)
and density contours (dashed curves) for
$M_{\rm a} = 10^{-5}\Msun$ and $\psi_{\rm a} = 0.1\psi_{*}$.
Coordinates measure altitude.
Density contours are drawn for $\eta\rho_{\rm max}$
($\rho_{\rm max} = 2.52\times 10^{14} {\rm \, g \, cm}^{-3}$),
with $\eta = 0.8,\, 0.6,\, 0.4,\, 0.2,\, 10^{-2},\, 10^{-3},\, 10^{-4},\,
10^{-5},\, 10^{-6},\, 10^{-12}$.
Convergence residuals are less than $10^{-3}$.
[From \citet{pay04}.]
(\emph{middle}) Contours of Lorentz force per unit volume for
the same $\eta$ values.
(\emph{bottom}) Contours of pressure gradients for the same $\eta$ values.
Note that colatitude $= 0^{\circ}$ at the pole.
}
\label{fig:polar}
\end{figure}

In the process of magnetic burial, material accreting onto a neutron star
accumulates in a column at the magnetic polar cap, until
the hydrostatic pressure at the base
of the column overcomes the magnetic tension and
matter spreads equatorward,
dragging along
frozen-in polar magnetic field lines to form
an equatorial magnetic belt or `tutu'.
Figure \ref{fig:polar}(a),
reproduced from \citet{pay04},
illustrates the equilibrium configuration
obtained for $\Ma = 10^{-5}\Msun$,
where $\Ma$ is the total accreted mass.
The polar mountain of accreted material (dashed contours) and
the pinched, flaring, equatorial magnetic belt are evident
\citep{mel01,pay04}.
The equatorial magnetic field strength increases
in inverse proportion to the surface area of the
equatorial belt, by flux conservation.
The Lorentz force per unit volume
exerted by the compressed equatorial field
[Figure \ref{fig:polar}(b)]
balances the thermal pressure gradient [Figure \ref{fig:polar}(c)],
and gravity, preventing the accreted matter from spreading
to the equator.

In the steady state, the equations of ideal magnetohydrodynamics (MHD)
reduce to the force balance equation
(CGS units)
\begin{equation}
\nabla p + \rho\nabla\phi - {(4\pi)}^{-1}(\nabla\times {\bf B})\times {\bf B} = 0,
\label{eq:forcebalance}
\end{equation}
where ${\bf B}$, $\rho$,
$p = c_{\rm s}^2\rho$, and
$\phi(r) = GM_{*}r/R_{*}^{2}$
denote the magnetic field, fluid density,
pressure, and gravitational potential respectively,
$c_{\rm s}$ is the isothermal sound speed,
$M_{*}$ is the mass of the star, and
$R_{*}$ is the stellar radius.
In spherical polar coordinates $(r,\theta,\phi)$,
for an axisymmetric field
${\bf B} = \nabla\psi(r,\theta)/(r\sin\theta)\times\hat{\bf e}_\phi$,
equation (\ref{eq:forcebalance}) reduces to the Grad-Shafranov equation
\begin{equation} 
\Delta^2\psi = F^{\prime}(\psi)\exp[-(\phi-\phi_0)/c_{\rm s}^2],
\label{eq:gradshafranov}
\end{equation}
where
\begin{equation} 
\Delta^2 =  \frac{1}{\mu_0 r^2\sin^2\theta} \
\left[ \frac{\partial^2}{\partial r^2} + \
\frac{\sin\theta}{r^2}\frac{\partial}{\partial\theta}\left(\frac{1}{\sin\theta}\frac{\partial}{\partial\theta}\right) 
\right]
\label{gs}
\end{equation}
is the Grad-Shafranov operator,
$F(\psi)$ is an arbitrary function of the magnetic flux $\psi$,
and we set $\phi_{0} = \phi(R_{*})$.
In this paper, as in \citet{pay04},
we fix $F(\psi)$ uniquely by connecting 
the initial dipolar state (with $\Ma = 0$) and final distorted state
[e.g. Figure \ref{fig:polar}(a)]
via the integral form of the flux freezing
condition of ideal MHD, viz.
\begin{equation} 
\frac{dM}{d\psi} = 2\pi\int_C \frac{ds \, \rho}{|\vv{B}|}.
\label{eq:fpsi}
\end{equation}
Here, $C$ is any magnetic field line, and the mass-flux distribution
$dM/d\psi$ is chosen to capture the magnetospheric geometry,
e.g. the accretion stream is funneled magnetically onto the pole,
with
$dM/d\psi \propto \exp(-\psi/\psi_{\rm a})$,
where  $\psi_{\rm a}$
is the polar flux.
We also assume north-south symmetry
and adopt the boundary conditions
$\psi =$ dipole
at $r = R_{*}$ (line tying),
$\psi = 0$ at $\theta = 0$,
and $\partial\psi/\partial r = 0$ at large $r$.
The line-tying approximation artificially prevents
accreted material from sinking,
so the computed $\rho$ is an upper limit.
Equations (2) and (3)
are solved numerically using an iterative relaxation scheme
and analytically by Green functions,
producing equilibria like Figure \ref{fig:polar}.

\subsection{Equatorial magnetic belt}
\label{sec:equatorialbelt}
\begin{figure}
\plotone{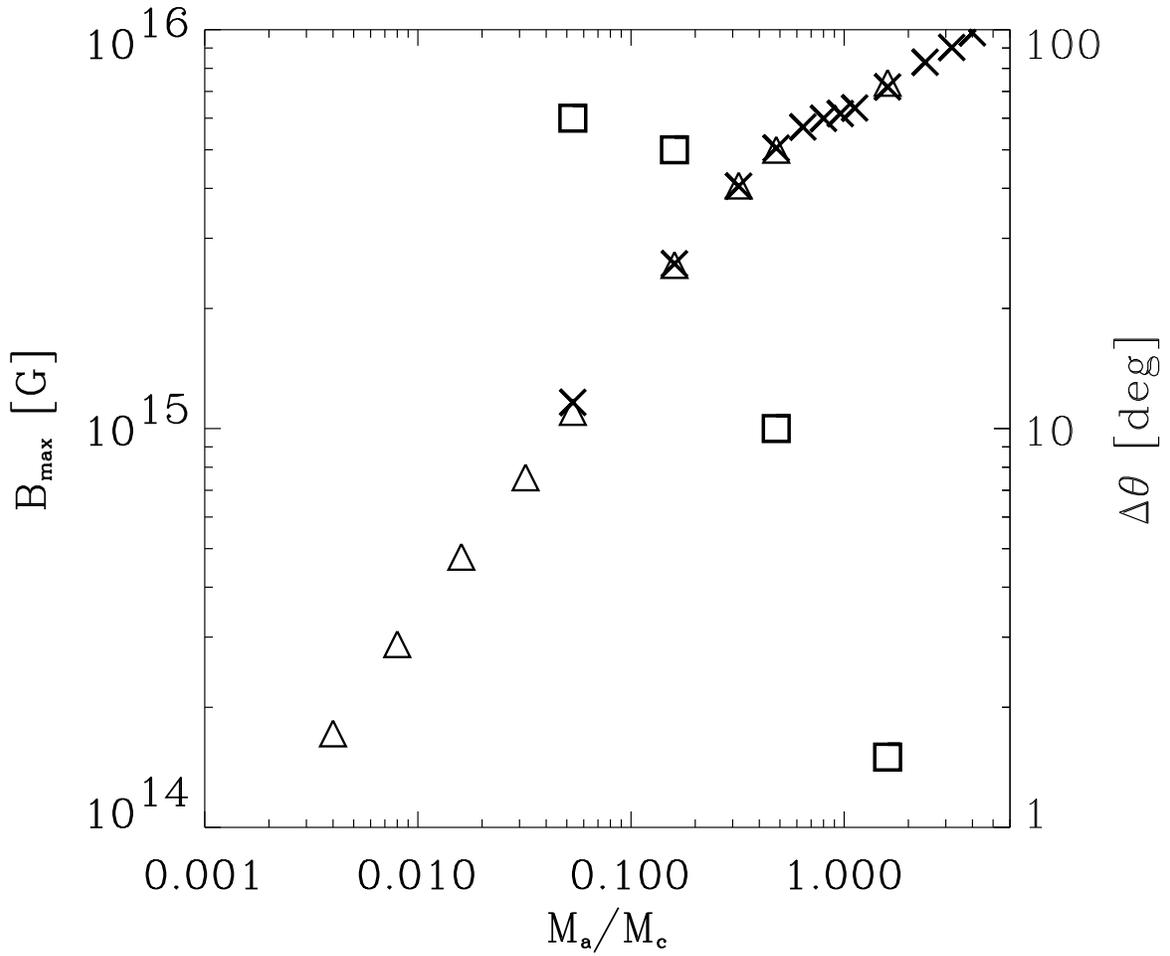}
\caption{Maximum magnetic field
in the equatorial belt, $B_{\rm max}$,
computed numerically for
$h/R_{*} = 2 \times 10^{-2}$ (\emph{crosses})
and $h/R_{*} = 5\times 10^{-5}$ (\emph{triangles})
and scaled using
$B_{\rm max}\propto h^{-1}$,
plotted together with the
half-width half-maximum thickness, $\Delta\theta$
(\emph{squares}), as a function of accreted mass, $\Ma$.
}
\label{fig:magma}
\end{figure}

The compressed magnetic field in the equatorial belt emerges
approximately perpendicular to the stellar surface, with opposite
sign in the two hemispheres.
Near the equator,
$B$ falls off roughly as
$\exp[-0.7(\pi/2-\theta)/\Delta\theta] \exp[-(r-R_{*})/h]$, where
$h = c_{\rm s}^{2} R_{*}/GM_{*}$
is the hydrostatic scale height,
$\Delta\theta$ is the belt thickness, and the factor 0.7 comes from
empirically fitting to the numerical results.
The maximum magnetic field strength
in the belt, $B_{\rm max}$,
computed numerically as a function of $\Ma$,
is plotted in
Figure \ref{fig:magma}.
It is fitted by
\begin{equation}
B_{\rm max} =
\cases{
2.0\times 10^{16}(\Ma/\Mc)^{0.91\pm 0.06}{\rm G}\quad \Ma\lesssim 0.4\Mc & \cr
6.3\times 10^{15}(\Ma/\Mc)^{0.32\pm 0.01}{\rm G}\quad \Ma\gtrsim 0.4\Mc & \cr
\label{eq:bmax}
}
\end{equation}
with
$\Mc = G M_{*} B_{0}^{2} R_{*}^{2}/8 c_{\rm s}^{4}
\sim 10^{-4}\Msun$,
where $B_{0}$ is the polar magnetic field strength
prior to accretion.
The scaling (\ref{eq:bmax}) agrees
with analytic theory for
small $\Ma$
\citep{pay04}.
Also plotted in Figure \ref{fig:magma} is the half-width
half-maximum thickness of the belt, fitted by
$\Delta\theta = 3^{\circ}(\Ma/\Mc)^{-1.5\pm 0.03}$.
We find that $\Delta\theta$ decreases as $B_{\rm max}$ increases,
as expected from magnetic flux conservation,
but not exactly as
$\Delta\theta \propto B_{\rm max}^{-1}$, because
$B_{r}$
is underestimated numerically
at the equator by $\sim 10$ per cent
(flux loss due to finite grid resolution).
Note that equation (\ref{eq:bmax}) and the above scalings of
$\Delta\theta$ versus $\Ma$ do not include Ohmic diffusion,
which is discussed further in \S 2.3 and \S \ref{discussion}.

Grad-Shafranov equilibria are difficult to compute directly
from (\ref{eq:gradshafranov}) and (\ref{eq:fpsi})
for $\Ma \gtrsim 1.6\Mc$,
because the magnetic topology changes abruptly:
magnetic bubbles form which are
disconnected from the surface, hinting at time-dependent
processes which the steady-state theory cannot describe.
In addition, the iterative relaxation scheme struggles
to handle steep field gradients.
Therefore the results in Figure \ref{fig:magma} for
$\Ma \geq 1.6\Mc$ are computed by direct numerical simulation using
ZEUS-3D, a multipurpose,
time-dependent, ideal-MHD code for astrophysical fluid dynamics
which uses staggered-mesh
finite differencing and operator splitting
in three dimensions
\citep{sto92}.
We load the Grad-Shafranov equilibrium for $\Ma = 1.6\Mc$
into ZEUS-3D;
double $\Ma$ quasistatically over 250 Alfv\'en times
with inflow (and hence radial ${\bf B}$)
boundary conditions;
stop the inflow and allow $\vv{B}$ to relax to a dipole at large $r$;
then iterate
to reach
$\Ma \sim 10^{-3}\Msun$
\citep{pay06}.
An isothermal equation of state is chosen
and self-gravity is switched off.
The experiment is performed for 
$h/R_{*} =  2\times 10^{-2}$ 
(for computational efficiency) and 
scaled to neutron star conditions ($h/R_{*} = 5\times 10^{-5}$)
according to $B_{\rm max}\propto R_{*}/h$;
this scaling is verified numerically for the range of $\Ma$ in
Figure \ref{fig:magma}.

\subsection{Stability and Ohmic relaxation}
\begin{figure}
\plotone{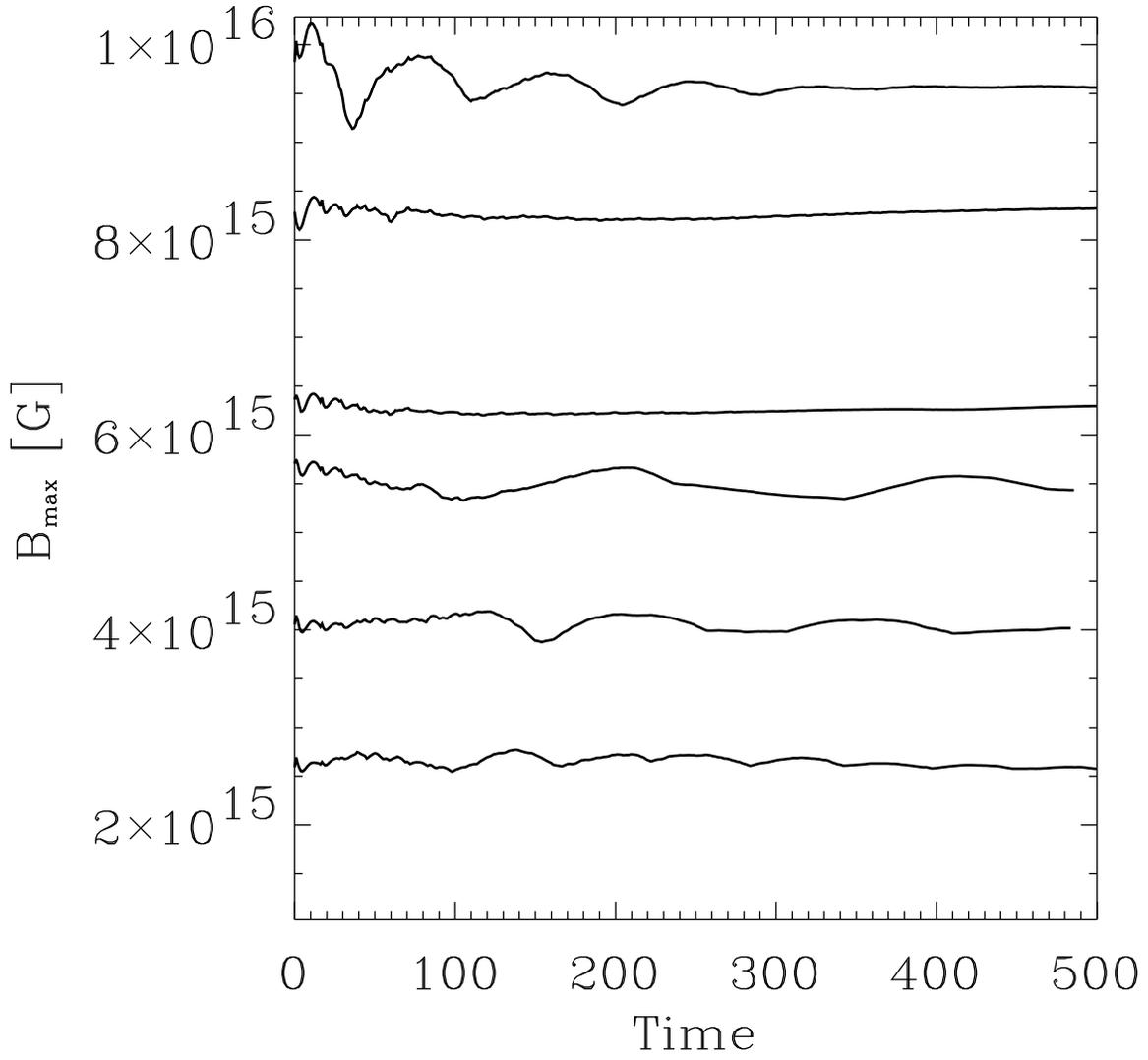}
\caption{
Maximum magnetic field strength, $B_{\rm max}$, 
as a function of time (in units of the Alfv\'en time)
for $\Ma/\Mc = 0.16,\, 0.32,\, 0.64,\, 1.12,\, 2.4,\, 4.0$ 
(\emph{bottom} to \emph{top}) when the equilibrium in Figure \ref{fig:polar} is
loaded into ZEUS-3D and perturbed slightly.
The configuration is stable.
}
\label{fig:magstable}
\end{figure}

Distorted ideal-MHD
equilibria are often disrupted by Parker and interchange instabilities.
Remarkably, however, this is not true for the equilibrium in
Figure \ref{fig:polar}.
Figure \ref{fig:magstable} shows the results of an experiment in which
the equilibrium is loaded into ZEUS-3D, perturbed,
and evolved for 500 Alfv\'en times
\citep{pay06}.
[The Alfv\'en time is defined as $h/v_{\rm A}$, where $v_{\rm A}$
is the Alfv\'en speed averaged over the grid.]
It is marginally stable; $B_{\rm max}$ oscillates via
magnetosonic (phase speed $\approx c_{\rm s}$) and Alfv\'en
(phase speed $\approx v_{\rm A}$)
modes which are
damped by numerical dissipation.
The configuration
\emph{already} represents the end point
(reached quasistatically)
of the nonlinear Parker instability of an initially dipolar
field overlaid with accreted material.
It is not interchange unstable because
line tying prevents flux tubes from squeezing past each other. 
That said, we caution that
we have not yet investigated the full gamut of 
three-dimensional MHD instabilities;
cross-field mass transport cannot be categorically ruled out.

Another pathway to cross-field
mass transport is if the magnetic belt relaxes by Ohmic diffusion,
either during quiescence or during a burst itself.
During quiescence, diffusion occurs most rapidly at the base of the
accreted layer.
In the relaxation time approximation, with
Coulomb logarithms set to $10$, the electron-phonon conductivity
for a hydrogen-helium mixture is 
$\sigma = 6.3\times 10^{24}(\rho/10^{11}{\rm g \, cm}^{-3}){\rm \, s}^{-1}$,
the density at the base of the accreted layer is
$\rho=6.2\times 10^{10}(\Ma/10^{-5}\Msun)^{3/4}{\rm \, g \, cm}^{-3}$
\citep{bro98},
and hence the Ohmic diffusion time-scale across the equatorial belt
(at the base of the accreted layer) is
$t_{\rm d}=
4\pi\sigma R_{*}^{2}\Delta\theta^2/c^2 =
2.6\times 10^{7}
(\Ma/10^{-5}\Msun)^{-9/4}
({T}/{10^{8}{\rm \, K}})^{-1}
{\rm yr}$,
much longer than the burst time-scale,
where we rewrite
$\Delta\theta$ in terms of $\Ma$.
During a burst, diffusion occurs most rapidly in the burning layer.
The temperature of the burning layer
rises isobarically until the radiation pressure
dominates the hydrostatic pressure, reaching
$T \sim 10^{8}$ K for a typical ignition column \citep{bro04}.
The elevated temperature and reduced density lower
$\sigma$ and accelerate Ohmic diffusion. 
In the burning layer ($\rho \approx 10^{6}$ g cm$^{-3}$),
we find $t_{\rm d} = 1.5\times 10^{3}$ yr,
still much longer than the burst time-scale.

Ohmic diffusion will be modelled self-consistently in a future paper.

\section{Thermal transport across the equator}
Does the equatorial magnetic belt impede thermal transport enough
to stop the thermonuclear flame
in a type I X-ray burst from spreading from one hemisphere to another?
A burst is initiated locally by a thin shell thermal instability
\citep{sch65}.
As the nuclear burning time-scale is much shorter than the
time to accrete the minimum column for ignition,
the accreted layer ignites at a single point,
most likely at the equator where gravity is rotationally reduced,
and the thermonuclear flame spreads away either as a deflagration front
\citep{fry82,bil95},
by detonation
\citep{fry82,zin01},
or as a cyclone driven by zonal shear
\citep{spi02}.
Detonation, which occurs when the nuclear burning time-scale is
less than the vertical sound crossing time-scale, requires
a thick ($\sim 100$ m) column of fuel
and hence a low accretion rate
($\lesssim 10^{-11.5}\Msun$ yr$^{-1}$).
Deflagration
occurs most commonly, with the front propagating at a speed set by the
heat flux and convection
\citep{fry82}.
Note that
temperature fluctuations leading to single-point ignition are difficult
to create because the
sound crossing time around the star is very short, but
Coriolis forces can assist by balancing sideways pressure gradients when
the star is rapidly rotating \citep{spi02}.

We estimate below
how the equatorial magnetic belt modifies
heat transport
in these scenarios.

\subsection{Conduction}
The thermal conductivity $\kappa$ of degenerate
electrons at the atmosphere-ocean boundary of a magnetized neutron star,
where H/He burning occurs,
was calculated by \citet{pot99}.
We extract
approximate values for the conductivity perpendicular to
the magnetic field from
Figures 5 and 6 of that paper, obtaining
$\kappa_{\perp} \approx 10^{7.5} (\rho/10^{6}{\rm g \, cm}^{-3})\
(T/10^{8}{\rm K})^{2}(B/10^{15}{\rm G})^{-2}$
in units of
${\rm erg \,\, s}^{-1}{\rm cm}^{-1}{\rm K}^{-1}$
in the regime
$10^{4}$ g cm$^{-3}\lesssim\rho\lesssim 10^{9}$ g cm$^{-3}$, 
$10^{12}$ G $\lesssim B \lesssim 10^{15}$ G, and
$10^{6}$ K $\lesssim T \lesssim 10^{8}$ K.
Note that $\kappa_{\perp}$
at the equator
is reduced $10^{6}$ times relative to its value
before accretion commences
($\propto B^{-2}$)
and, for reference, one has
$\kappa_{\perp}/\kappa_{\parallel}\sim 10^{-5}$
at $T = 10^{8}$ K,
$B = 10^{12}$ G
\citep{pot99}.
We estimate the conduction time-scale across the magnetic barrier from
$t_{\rm cond} =R_{*}^{2}\Delta\theta^{2}C \rho/\kappa_{\perp}$,
where
$C
= 10^{4}(T/10^{8}{\rm K}){\rm \, erg \,\,  g}^{-1}{\rm \, K}^{-1}$
is the specific heat capacity of a degenerate Fermi gas
\citep{bro98},
finding
\begin{equation}
t_{\rm cond}= \
27
({\Ma}/{\Mc})^{-3}
({T}/{10^{8}{\rm K}})^{-1}
({B}/{10^{15}{\rm G}})^{2}
{\rm \, yr} \, ,
\end{equation}
which is safely longer than the duration of the burst.
Note that this estimate does not include corrections due to electron-electron scattering
and impurity scattering.
Furthermore, our fit smoothes over the
wiggles in the solid curves of Figures 5 and 6 in \citet{pot99},
which arise from quantization into Landau orbitals.

\subsection{Radiation}
\label{sec:radiation}
In a strong magnetic field,
photons polarized perpendicular to $\vv{B}$
dominate radiative transport.
In the burning (helium) layer,
where Thomson scattering dominates,
the Rosseland mean opacity $\perp \vv{B}$ is given by 
$\kappa_{\rm R} = 1.3\times 10^{-6}
({T}/{10^8 {\rm K}})^{2}({B}/{10^{15}{\rm G}})^{-2}
{\rm g}^{-1}
{\rm \, cm}^{2}$,
decreasing in inverse proportion to the square of the cyclotron frequency
($\omega_{\rm c}$) \citep{jos80,fry82}.  Hence
the optical depth of the barrier is
$\tau_{\rm R} \
= \kappa_{\rm R}\rho R_{*}\Delta\theta$, i.e.
\begin{equation}
\tau_{\rm R} \
= 6.8\times 10^{4} \
(\Ma/\Mc)^{-3/2} (T/10^{8}{\rm K})^{2}(B/10^{15}{\rm G})^{-2}.
\label{eq:opdepth}
\end{equation}
The optical depth (\ref{eq:opdepth}) is a \emph{lower} limit
obtained by assuming that radiation is transported
from one hemisphere to the other near the surface,
at the depth of the burning layer
($\rho\sim 10^{6}{\rm g \, cm}^{-3}$)
rather than at the base of the accreted column
($\rho\sim 10^{11}{\rm g \, cm}^{-3}$).
Hence the peak flux penetrating the barrier is
$F_{\rm burst} = L_{\rm burst}e^{-\tau_{\rm R}}/4\pi R_{*}^{2}
= 8\times 10^{24} e^{-\tau_{\rm R}}$
erg \, s$^{-1}$ \, cm$^{-2}$
(for $L_{\rm burst} = 10^{38}{\rm erg \, s}^{-1}$),
which is insufficient to heat
the other (quiescent) hemisphere
above its ignition temperature
($\sim 10^{8}$ K)
(assuming thermal equilibrium
and
applying the Stefan-Boltzmann law).
Interestingly,
once $B_{\rm max}$ exceeds $10^{17}$ G, we find $\tau_{\rm R} \lesssim 1$ and
the magnetic belt becomes optically thin.
Vertical heat propagation is not considered here.

Equation (\ref{eq:opdepth}) can be generalized in several ways.
Mode coupling \citep{mil95}
can cause a fraction $\sim 0.2\omega/\omega_{\rm c}$ of the
perpendicular mode photons to convert into parallel mode photons,
so that the opacity scales $\propto B^{-1}$,
increasing the field strength required for the belt to become
optically thin.
Vacuum polarization \citep{oze03}, along with proton-cyclotron resonance,
gives sharp spikes in the
frequency response of the atmospheric opacity.
However, the resonant densities for vacuum polarization
($\rho\sim 10^{-3}$ \, g \, cm$^{-3}$)
are achieved at shallow depths, well above the burning layer.

\subsection{Convection}
\citet{bil95} estimated the convective speed
in the burning layer in
terms of the mixing length, $l_{\rm m}$, and thermal time-scale,
$t_{\rm th}$, finding
$v_{\rm c} \approx c_{\rm s}(l_{\rm m}/h)^{1/3}(c_{\rm s}R_{*}^{2}/G M_{*} t_{\rm th})^{1/3}
\approx 10^{6}{\rm \, cm\,\, s}^{-1}$.
This is consistent with the
upper bound
$10^{7}{\rm cm \, s}^{-1}$
obtained if
$L_{\rm burst}$ is transported entirely by the mechanical flux $\rho v_{\rm c}^{3}$.
It is also consistent with the observed burst rise time
$R_{*}/v_{\rm c} \lesssim 1$ s
\citep{spi02}.
Convection is stabilized magnetically if the
magnetic tension exceeds the ram pressure
$\rho v_{\rm c}^{2}$, which occurs for
$B_{\rm max} \gtrsim (8\pi\rho v_{\rm c}^{2})^{1/2}\approx 5\times 10^{9}$ G.
This condition is met comfortably in the equatorial belt.
The magnetic field can also quench convection
\citep{gou66}.

\subsection{Ageostrophic shear flow}
\citet{spi02}
suggested that
inhomogeneous cooling drives zonal currents which are unstable to the
formation of cyclones, as
in planetary atmospheres.
This may explain why the coherent oscillations
observed in the tails of some type I X-ray bursts
persist for many rise times.
The drift in oscillation frequency ($\sim$ Hz)
during the burst is attributed to the Coriolis drift
of the cyclone in the frame of the star,
although theory predicts larger frequency drifts ($\sim 10$ Hz)
than those observed.
The oscillation amplitude in the burst tails ($\sim 10 \%$)
is governed by processes other than magnetic burial, e.g.
small-scale magnetic fields generated by an MHD dynamo in the
burning front, which are
confined to the ashes after the burst,
while the freshly accreted matter remains unmagnetized
\citep{spi02}.

Ageostrophic shear flow
moves hot material ahead
of the burning front and draws fresh fuel into it
at the flame speed
$v_{\rm flame}\approx
2\times 10^{7}{\rm \, cm\,\, s}^{-1}(h_{\rm hot}/10^{3}{\rm cm})
(f/0.32 {\rm kHz})^{-1}(t_{\rm nuc}/0.1 {\rm \, s})^{-1}$,
where
$h_{\rm hot}$ is the scale height of the incinerated ocean,
$f = 2\Omega\cos\theta$ is the local Coriolis parameter,
$\Omega$ is the angular frequency of the star,
and
$t_{\rm nuc}$ is the nuclear burning time-scale.
Magnetic tension stabilizes ageostrophic shear for
$B \gtrsim (8\pi\rho v_{\rm flame}^{2})^{1/2} \approx 4.5\times 10^{10}$ G
$\ll B_{\rm max}$.
In other words,
when the cyclonic flame runs into the equatorial magnetic belt,
it is reflected;
conversely, ageostrophic shear cannot
disrupt the belt.
Latitudininal shear instabilities, a possible source of burst
oscillations \citep{cum05}, may disrupt the magnetic belt
and provide one motivation for extending our burial model to
three dimensions in the future.

\section{DISCUSSION}
\label{discussion}
Polar magnetic burial
creates
an intense, equatorial belt of magnetic field which can
thermally isolate the
magnetic hemispheres of an accreting neutron star.
The maximum magnetic field strength in the belt,
$B_{\rm max} = 5.6\times 10^{15} (\Ma/10^{-4}\Msun)^{0.32}$ G,
is sufficient to prevent heat transport by
conduction, radiation, convection,
and ageostrophic shear.
The conduction time-scale
$t_{\rm cond}\sim 27$ yr
exceeds the cooling time of the incinerated material;
the magnetic belt is opaque
(optical depth $\sim 7\times 10^{4}$);
convection is stabilized magnetically 
($\rho v_{\rm c}^{2} \ll B_{\rm max}^{2}/8\pi$);
and so is ageostrophic shear
($\rho v_{\rm flame}^{2} \ll B_{\rm max}^{2}/8\pi$).
However, the conclusion that the hemispheres are thermally isolated
by the magnetic belt is less secure at large accreted masses
($\Ma \sim 0.1\Msun$), where the Grad-Shafranov and ZEUS-3D calculations
in section 2 are hampered by numerical difficulties, and Ohmic diffusion
(which we do not incorporate self-consistently) becomes important.

Thermal isolation of the magnetic hemispheres is consistent
with the highly sinusoidal light curves of
millisecond oscillations in thermonuclear bursts in LMXBs
\citep{mun02}.
During the rise of the burst, oscillations
are caused by the spreading of a hot spot, probably in the form of
a cyclone \citep{spi02}.
In the tail of the burst, the hot spot is sequestered in one hemisphere
\citep{mun02}, and
misalignment of the magnetic and spin axes guarantees that
we observe persistent oscillations at
the spin frequency.
Furthermore, it is observed that
bursts occasionally occur in quick succession, separated by a
$5 - 10$ minute interval,
compared to an interval of several hours between typical bursts
\citep{lew93}.
Such burst pairs are consistent with the model of magnetic burial:
if a burst ignites one hemisphere, the other remains dormant
and can, in principle, ignite shortly thereafter
when more material accretes.
If this explanation is valid, one would expect that there are no triple
bursts, that the fluences emitted in each half of a burst pair
are roughly equal, and that the number of burst pairs relative
to normal bursts increases as
the ignition probability per unit time increases (relative to
$\dot{M}_{\rm a}$).\footnote{
If the ignition probability is higher (given a specific $\dot{M}_{\rm a}$),
it is more likely that the second hemisphere will ignite shortly after
the first (before the first has a chance to refill and ignite again).
A faster burst rate may be associated with more pairs,
but gaps in the satellite data make it difficult to be sure.
}
A more detailed calculation is required to resolve whether
the 10-min intervals are caused by delayed heat propagation
through the magnetic belt.

\citet{mun02}
showed that the bright burning region
must cover $80^{\circ} - 110^{\circ}$ in latitude
in order to match the upper limit on the observed ratio of harmonic
to fundamental amplitudes,
which implies $\Delta\theta \lesssim 10^{\circ}$.
If the scaling of $\Delta\theta$ versus $\Ma$ in \S \ref{sec:equatorialbelt}
is taken at face value,
this requires $\Ma > 0.5\Mc$.
Furthermore, for the belt to be optically thick, we need
$\Ma \lesssim 2\Mc$ (\S \ref{sec:radiation}).
Apparently then, only a narrow range of accreted
masses ($0.5\lesssim \Ma/\Mc \lesssim 2$) can account for the
observations.
However, the above $\Delta\theta$ scaling,
which depends steeply on $\Ma$,
does not tell the whole story, because
Ohmic diffusion is not incorporated self-consistently.
Equatorward hydromagnetic spreading
is arrested when the accretion time-scale exceeds the
Ohmic diffusion time-scale \citep{bro98},
softening the dependence of $B_{\rm max}$ and $\Delta\theta$
on $\Ma$
in the manner described by \citet{mel05}.
Ohmic diffusion arrests magnetic compression
for $\Ma > M_{\rm d}$,
where
$M_{\rm d} = 3.4\times 10^{-7}(T/10^{8} {\rm K})^{-2.2}(\dot{M}_{\rm a}/10^{-8}\Msun {\rm yr}^{-1})^{0.44} \Msun$ \citep{mel05} is the accreted mass
at which the accretion time-scale exceeds the Ohmic diffusion
time-scale.
Including this effect, $\Delta\theta$
is modified to the maximum of
$3^{\circ} (\Ma/10^{-4}\Msun)^{-1.5}$ and
$3^{\circ}(\Ma/10^{-4}\Msun)^{0.5}(\dot{M}_{\rm a}/10^{-8}\Msun {\rm yr}^{-1})^{-0.5}]$,
and the belt remains optically thick even for
$\Ma \gtrsim 2\Mc$ (\S \ref{sec:radiation}).

Only 70 out of 160 LMXBs undergo bursts, and only
13 exhibit millisecond oscillations.
Does the equatorial belt model respect these statistics?
If accretion occurs at the Eddington rate,
$\dot{M}_{\rm a} \approx 10^{-8}\Msun {\rm yr}^{-1}$,
it takes less than $10^{4}$ yr to achieve $\Ma > \Mc$
and screen the polar magnetic field, allowing bursts to ignite
and creating a thermally insulating equatorial belt.
However,
for $\Ma\gtrsim 2\times 10^{3}\Mc$,
the belt becomes optically thin to X-rays
and
$t_{\rm cond}$ drops below the duration of the burst,
potentially allowing the flame to engulf the entire star
in LMXBs older than a few
times $10^{7}$ yr.
If accretion occurs at
$\dot{M}_{\rm a} \sim 10^{-11}\Msun {\rm yr}^{-1}$, as in
accreting millisecond pulsars
\citep{cha03},
it takes $10^{7}$ yr to achieve
$\Ma > \Mc$, screen the polar magnetic field,
and allow bursts to ignite.
Therefore, we do not expect bursts from all accreting millisecond pulsars,
but when bursts do occur, we expect to detect oscillations at some level
because we have $\tau_{\rm R} \gg 1$ and $t_{\rm cond} \gg$
(burst time-scale) for $\Ma \lesssim 10^{-3}\Msun$.
Such is the case, within current observational sensitivity,
for SAX J1808.4-3658 and XTE J1814-338.
Sources like
XTE J1814-338, whose oscillations contain
harmonics exceeding
$25$ per cent of the peak amplitude
\citep{str03b}, may have semi-transparent equatorial belts;
perhaps this millisecond pulsar has experienced interruptions
in its accretion history resulting in  $\Ma < \Mc$.
Note that, for $\dot{M}_{\rm a}\lesssim 10^{-10}\Msun {\rm \, yr}^{-1}$,
Ohmic diffusion may magnetize the freshly accreted material
\citep{cum01},
increasing the polar magnetic
field strength and suppressing bursts.
Also, ignition is more likely near the equator,
where gravity is rotationally reduced,
but this is where the magnetic field is strongest.
The issues of conservative mass-transfer \citep{tau00} and
differential rotation leading to shear instabilities \citep{cum05}
are not considered here.

On the face of it, high-mass X-ray binaries (HMXBs)
accrete many times $\Mc$,
yet a strong magnetic field survives.  However, one needs to be cautious.
While the accretion rate in HMXBs can reach $10^{-10}$--$10^{-8} \Msun {\rm yr}^{-1}$
during either atmospheric Roche lobe overflow (close binaries with orbital period $< 5$ days)
or stellar wind accretion (wider binaries),
the typical lifetime of HMXBs as strong X-ray sources is $10^4$--$10^5$ yr
\citep{urp98}.
Sub-Eddington accretion rates ($10^{-13}$--$10^{-10} \Msun {\rm yr}^{-1}$)
characterize the rest of the companion's
main-sequence evolution for $10^6$--$10^7$ yr.
These scenarios yield roughly $\Ma \sim \Mc$.
Mass transfer can also be nonconservative, further reducing $\Ma$, e.g.,
in intermediate-mass X-ray binaries \citep{tau00}.
Of course, despite these cautionary remarks, it may well
be that some HMXBs do accrete many times $\Mc$.
If so, then either HMXBs are a counterexample to the simple
magnetic burial model we have calculated,
or else cross-field transport by Ohmic diffusion
(which we do not incorporate self-consistently)
becomes important.

A recent model for doubly peaked bursts which are too weak
to cause photospheric expansion
\citep{bha06a}
proposes that a burning front forms quickly
after ignition at or near a pole and propagates quickly towards
the equator where it stalls.  After a delay, the burning front
speeds up again into the opposite hemisphere.
While the time between burst peaks ($\sim 10$ seconds) is less than
that estimated for burst propagation across the magnetic belt,
a weak magnetic belt may explain why the burning front stalls
at the equator.
 
Numerical difficulties associated with steep field gradients
prevent us from verifying
whether our model scales up to $\Ma \sim 0.1\Msun$.
However, if we continue to respect mass-flux conservation
as in \S \ref{sec:gradshaf},
it is probable that accreting extra mass does not eliminate
the magnetic barrier. Even if, for example,
there are instabilities which
disrupt the equatorial magnetic belt above $\sim 10^{-4}\Msun$,
the belt does not stay disrupted; magnetic burial ensures that
it reforms as soon as a further $\sim 10^{-5}\Msun$ is accreted \citep{pay04}.
Given that $0.1\Msun$ is typically accreted
in LMXBs, the chances of catching the belt in its disrupted state
are slim.


\acknowledgments
{We thank Duncan Galloway for pointing out to us that sinusoidal
light curves are a signature of hemispheric emission,
and alerting us to the existence of burst pairs.
We thank an anonymous referee for pointing out to us that the burial
model may imply the existence of pairs of bursts in quick succession,
and for comments which improved the treatment of the crust and burning
physics in the manuscript.
This research was supported in part by an
Australian Postgraduate Award.
}



\clearpage

\clearpage

\clearpage

\end{document}